\documentclass[9pt,twocolumn,twoside]{osajnl}
\journal{ol} 
\setboolean{shortarticle}{true}
\title{Generation of 14.0W of single frequency light at 770 nm by intracavity frequency doubling}

\author[1,*]{Minho Kwon}
\author[1,2]{Peiyu Yang}
\author[1]{Preston Huft}
\author[1]{Christopher Young}
\author[1]{Matthew Ebert}
\author[1]{Mark Saffman}

\affil[1]{Department of Physics, University of Wisconsin-Madison, 1150 University Ave., Madison, WI 53706, USA}
\affil[2]{Quantum Institute for Light and Atoms, School of Physics and Material Science, East China Normal University, Shanghai 200062, China}

\affil[*]{Corresponding author: mkwon22@wisc.edu}
\dates{\today}

\doi{}

\begin{abstract}
We present a continuous, narrow-linewidth, tunable laser system that outputs up to 14.0 W at 770 nm.
The light is generated by frequency doubling 18.8 W of light from a 1540 nm fiber amplifier that is seeded by a single mode diode laser  achieving >74\% conversion efficiency.
We utilize a Lithium Triborate Crystal in an enhancement ring cavity.
The low intensity noise and narrow linewidth of the 770 nm output are suitable for cold atom experiments.
\end{abstract}

\setboolean{displaycopyright}{false}

\begin{document}

\maketitle

Near-infrared(NIR) lasers around 760 - 780 nm have many applications including laser cooling and trapping of atoms, atomic state manipulation, and spectroscopy of Oxygen(760.8, 763.8 nm)\cite{Greenblatt1990}, Potassium(766.7, 770.1 nm)\cite{Tiecke2010} and Rubidium(780.2 nm)\cite{Steck}.
Iodine \(\text{I}_{2}\) rovibrational lines\cite{Rakowsky1989} at 770.7 nm  provide excellent frequency references.
In the context of quantum computing, magic wavelengths for optical trapping of cold alkali atoms\cite{Li2019,Wang2012a} near this wavelength exist, and the light can be used to rapidly drive Raman transitions between Potassium and Rubidium hyperfine ground states.

Common diode lasers at these wavelengths are typically power limited up to \(\sim 100\) mW.
Semiconductor optical amplifiers can boost the power to \(\sim3\) W, although typically with a poor spatial profile that reduces the usable power when diffraction-limited performance is required.
Furthermore, amplified spontaneous emission generates incoherent frequency noise extending over tens of nm, which leads to significant reduction of atomic coherence when the light interacts with atoms.
For atom trapping applications at IR wavelengths, fiber lasers are typically used for their high optical power and low frequency noise\cite{Mazurenko2019}.

An alternative route to high power tunable sources for atomic spectroscopy is based on second harmonic generation (SHG) or sum frequency generation(SFG)\cite{Guo2015} of a high power fiber laser.
This approach benefits from the availability of tunable, single mode laser diode sources, and high power fiber amplifiers that utilize mature industrial technology in telecom wavelength bands.
A variety of nonlinear crystals have been explored for this application.
Early efforts utilized the high nonlinearity of periodically-poled Lithium Niobate(PPLN) to achieve high power output\cite{Thompson2003,Lienhart2007,Wilson2011,Rengelink2016,Sane2012a,Lichtman2015,Runcorn2017} without introducing an enhancement cavity. Examples include  11 W at 780 nm for Rb cooling\cite{Sane2012a}, 313 nm for Be ion cooling and quantum gates\cite{Wilson2011}, and 319.8 nm for metastable Helium trapping\cite{Rengelink2016}.
Other efforts have utilized Magnesium Oxide doped PPLN (PPMgO:LN) for Rb/Cs cooling\cite{Zhang2018a} and K cooling\cite{Stern2010}.
Periodically-poled Potassium Titanyl Phosphate(PPKTP) in a Fabry-Perot enhancement cavity was used to achieve 95\% conversion efficiency\cite{Ast2011} and an output power of  1.05 W at 775 nm.

Periodically-poled crystals are engineered to exhibit high optical non-linearity utilizing quasi-phase matching.
Typically a single or double pass interaction  is sufficient to reach high conversion efficiency with high power sources.
Despite their high optical nonlinearity, these crystals are prone to thermal effects\cite{Liao2004,Spiekermann2004,Louchev2005,Ricciardi2010} leading to photorefractive beam distortion.
Furthermore, nonlinear induced absorption\cite{Furukawa2001} can cause permanent photochromic damage to the crystal, imposing another challenge for high power cw operation.

Instead, we use Lithium triborate(LBO), which is known for its robustness at high power\cite{Meier2011} and wide transparency window.
One drawback is the much weaker optical nonlinearity (\(d_{32}=0.67~\text{pm/V}\), Type-I NCPM), compared to PPLN (\(d_{33}=25~\text{pm/V}\)) and PPKTP (\(d_{33}=15-17~\text{pm/V}\))\cite{Shoji1997}.
This necessitates the use of a pulsed laser or an enhancement cavity.
Cavity-assisted conversion at 1064 nm was demonstrated\cite{Cui2019,Meier2010}, leading to >100 W of continuous 532 nm light.
We report here on the first high power, high efficiency frequency doubling of a 1540 nm laser using an enhancement cavity and LBO crystal\cite{Kwon2019}.
The combination of tunability, low intensity and frequency noise, and excellent spatial beam quality make this laser source suitable for experiments with Rb and K atoms. 

\begin{figure}
    \centering
    \includegraphics[width=0.9\columnwidth]{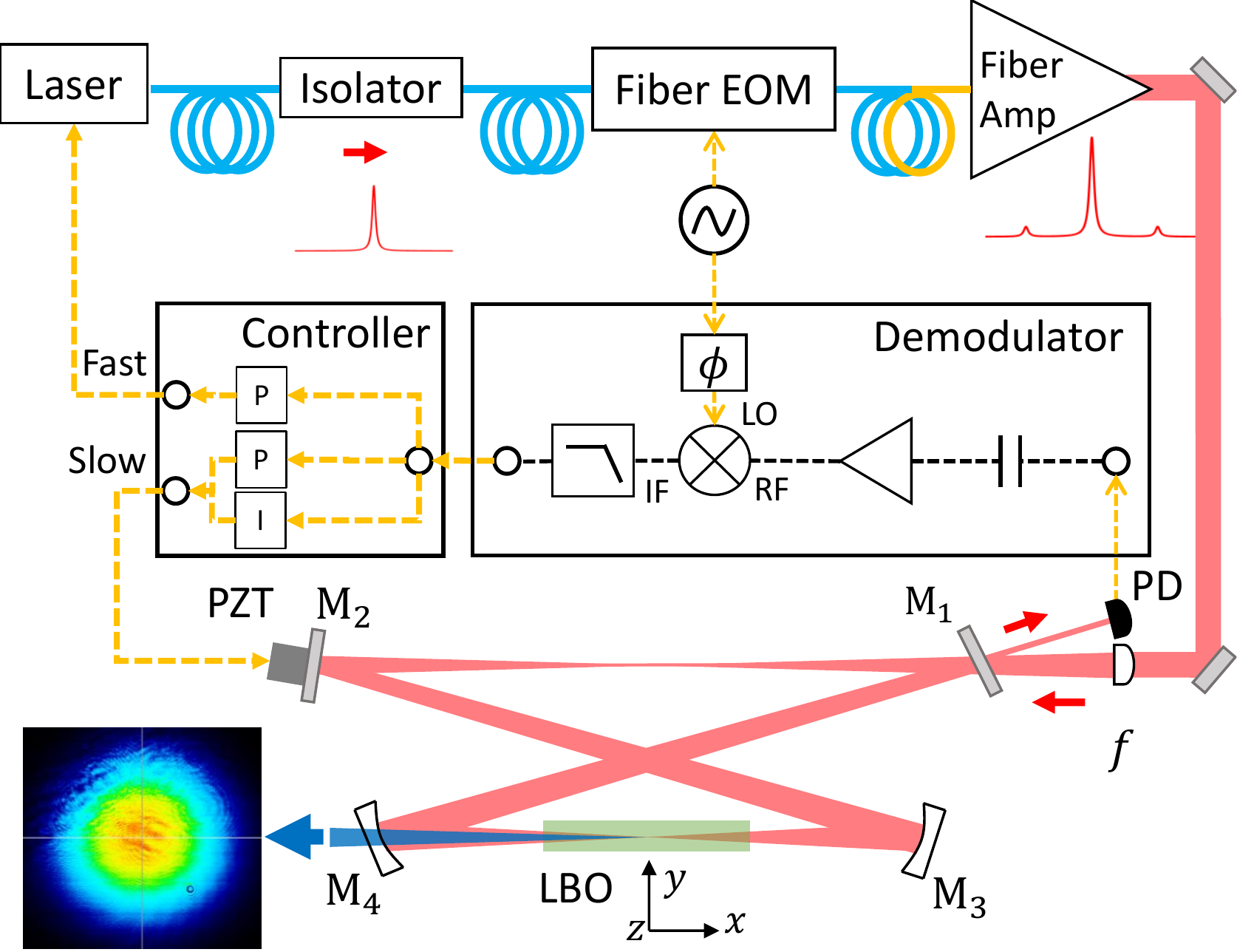}
    \caption{Setup for SHG of 1540 nm light. The beam is mode-matched  to the ring cavity with  a plano-convex lens (f=250 mm).
The cavity is comprised of two plano-concave mirrors(ROC=50 mm) and two flat mirrors in a bow-tie configuration, with angles of incidents(AOI) nominally \(8^{\circ}\).
Input coupler(\(\text{M}_{1}\)) is partially reflective at 1540 nm and the transmittance is optimized for impedance-matching.
The other cavity mirrors(\(\text{M}_{2}, \text{M}_{3}\) and \(\text{M}_{4}\)) are front-side HR coated at 1540 nm and rear-side AR coated for 1540 and 770 nm.
The cavity round trip length is 29.5 cm and the distance between the curved mirrors is 6.6 cm.}
    \label{fig:overall setup}
\end{figure}

A schematic of the setup is shown in Fig.\ref{fig:overall setup}.
A 20 W fiber amplifier (IPG Photonics EAR-20K-C-LP-SF) is seeded by a 1540 nm single frequency fiber laser (Redfern Integrated Optics RIO3135-3-46-1).
A fiber phase modulator (iXblue MPX-LN-0.1-P-P-FA-FA) is used for Pound-Drever-Hall\cite{Drever1983} locking of the SHG cavity.
The fiber amplifier output is linearly polarized with \(1/e^{2}\) beam diameter of 1.1 mm. The fundamental resonator mode is circular(\(w_{y,z}=40.6~\mu\)m) at the crystal center, which is also aligned to the midpoint between  the  concave mirrors.
AR coated LBO crystals cut at (\(\theta=90^{\circ}, \phi=0^{\circ}\)), with dimensions of \(3\times3\times30\) mm were purchased from United Crystals and UVisIR.

Type-I Non-Critical Phase Matching of LBO converts two \(z\)-polarized photons to a single \(y\)-polarized photon propagating along the \(x\)-axis.
The crystal is temperature-tuned to satisfy phase matching conditions, dictated by Sellmeier equations\cite{Kato1994}.
The observed single-pass temperature bandwidth of \(\Delta T\cdot l=8.68\pm 0.15^{\circ}\text{C}\cdot\text{cm}\) is consistent with reported values\cite{Kato1994}.
However, we find a phase-matching temperature of \(88^{\circ}\)C, which is \(22^{\circ}\)C lower than the predicted value(\(110^{\circ}\)C) \cite{Kato1994}.
The measurement was done with low power input such that the contribution from absorption induced heating was negligible.
We do not have an explanation for the discrepancy.
Heating becomes significant when operating the enhancement cavity with high power input, due to high circulating power and finite absorption.
It is therefore necessary to make small adjustments to the crystal oven temperature to achieve optimum output, dependent upon input power.
For the maximum input power of 18.8 W, optimum temperatures are about \(1.2^{\circ}\)C lower than that of the low-power input.
We employed two-stage active temperature control, where the base plate and the crystal oven are independently controlled by TECs.
This minimizes the impact on resonator path length while the crystal temperature is altered.
Furthermore, smaller thermal mass of the crystal and the oven allows faster thermal control.
Mechanical parts are all made from metal to minimize outgassing and degradation at high temperature operation.

The cavity finesse is \(\mathcal{F}_{\text{LP}}\sim 102\) at low circulating power and when frequency doubling is operating at the maximum input power the finesse is \(\mathcal{F}_{\text{HP}}\sim 59\), corresponding to a cavity linewidth of \(\kappa_{\text{HP}}/2\pi\sim 16\) MHz and build-up factor of \(\sim 19\).
The round-trip length of the cavity is adjusted with a ring piezo (Piezomechanik HPSt 150/14-10/12 HAg) attached to the flat mirror(\(\text{M}_{2}\)), providing up to \(12~\mu\)m stroke length.
The actuator range is particularly important for handling thermal effects from the crystal.
With a given stroke it can compensate a \(\sim 24~\mu\text{m}\) round-trip length change, which is enough to compensate \(>4^{\circ}\)C of crystal temperature variation.
The seed laser can be locked to the enhancement cavity for applications where intensity stability is more important than frequency excursion.
Our laser had an optical tuning range of \(\sim\)500 MHz with modulation bandwidth of tens of kHz.

We initially tried the H\"ansch-Couillaud\cite{Hansch1980a,Boon-Engering1997}(HC) technique to obtain an error signal for  cavity locking, as it does not require any modulation of the laser.
However the HC error signal could not withstand transient thermal effects at high power, and we switched to the PDH technique.
A detailed discussion of thermal effects is provided below.
Reflection from the input coupler(\(\text{M}_{1}\)) is sampled from an AR-coated window, and detected by a fast photodetector (Thorlabs DET08C).
The signal is amplified and demodulated to obtain a dispersive PDH error signal.
The chosen phase-modulation frequency of 130 MHz provides large capture range covering \(>25\%\) of the free spectral range(FSR = 0.96 GHz) of the cavity. 
Modulation depth is kept small such that the carrier has \(>97\%\) of the power.
This keeps the nonlinear process power-efficient, while minimizing intensity noise in the harmonic light caused by the sidebands, which are off-resonant and thus attenuated by >18 dB. 
\begin{figure}
    \centering
    \includegraphics[width=\columnwidth]{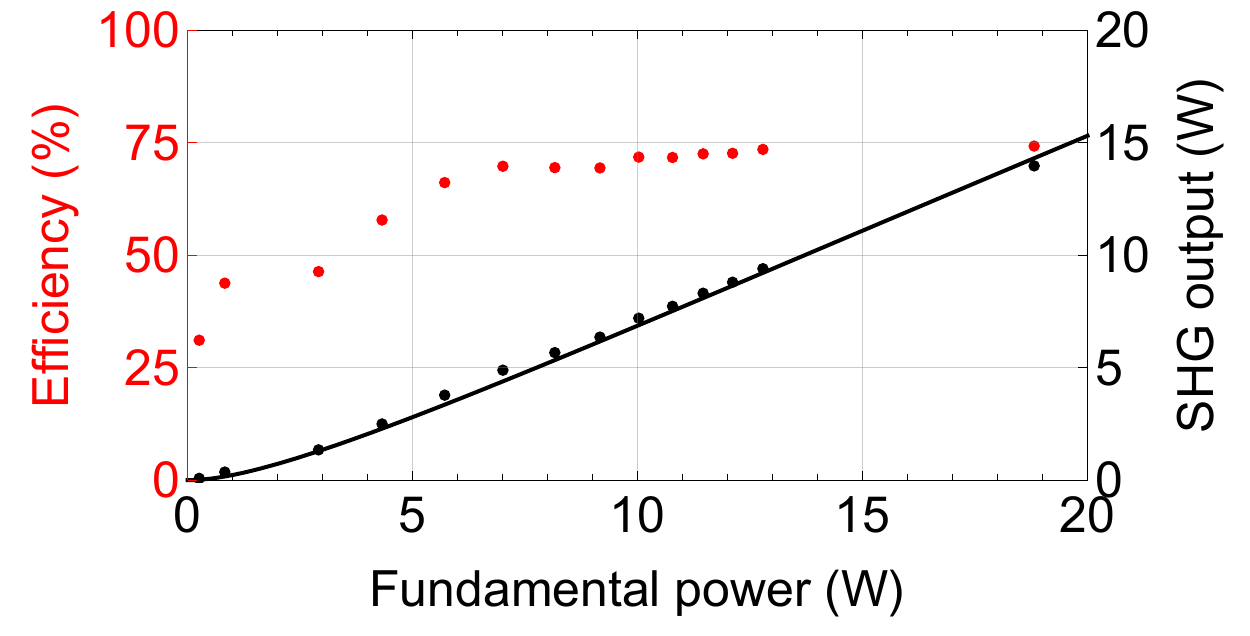}
    \caption{SHG output(Black points) and conversion efficiency(Red points) versus fundamental power. The black line is calculated SHG output based on measured optical parameters described in the text.}
    \label{fig:power throughput}
\end{figure}

A maximum harmonic output of 14.0 W at 18.8 W input power is observed as shown in Fig.  \ref{fig:power throughput}.
The conversion efficiency \(\epsilon=\text{P}_{2}/\text{P}_{1}\) saturates due to depletion of the circulating fundamental power and cavity loss.
Intracavity SHG output can be calculated\cite{Polzik1991} by solving the equation,
\begin{equation}
    \sqrt{\epsilon}=\frac{4T_{1}\sqrt{E_{NL}P_{m,1}}}{\left[2-\sqrt{1-T_{1}}(2-L-\sqrt{\epsilon E_{NL} P_{m,1}})\right]^{2}},
\end{equation}
where \(\text{T}_{1}\) is the transmission of the input coupler, \(L\) is the round-trip linear loss excluding the input coupler, \(E_{NL}\) is a single-pass nonlinear conversion coefficient, \(P_{m,1}=m P_{1}\) is the fundamental power coupled to the cavity, and \(m\) is the mode-matching coefficient to the \(\text{TEM}_{00}\) resonator mode.
We found \(\text{T}_{1}=5\%\) and \(E_{NL}=1.23\times10^{-6}~\text{W}^{-1}\) from direct measurements, \(L\sim1\%\) from the measured cavity finesse, and \(m\sim 0.95\) from the reflection dip.
Calculated SHG output is consistent with the parameters, as shown in Fig. \ref{fig:power throughput}.
The measured spatial profile of the SHG output had \(\text{M}^{2}<1.4\) and circular(\(1:0.97\)), as shown in Fig. \ref{fig:overall setup}.

Continuous wave operation requires active stabilization of the cavity length.
A servo-loop using an error signal generated by the PDH technique provided a robust lock.
\begin{figure}
    \centering
    \includegraphics[width=\columnwidth]{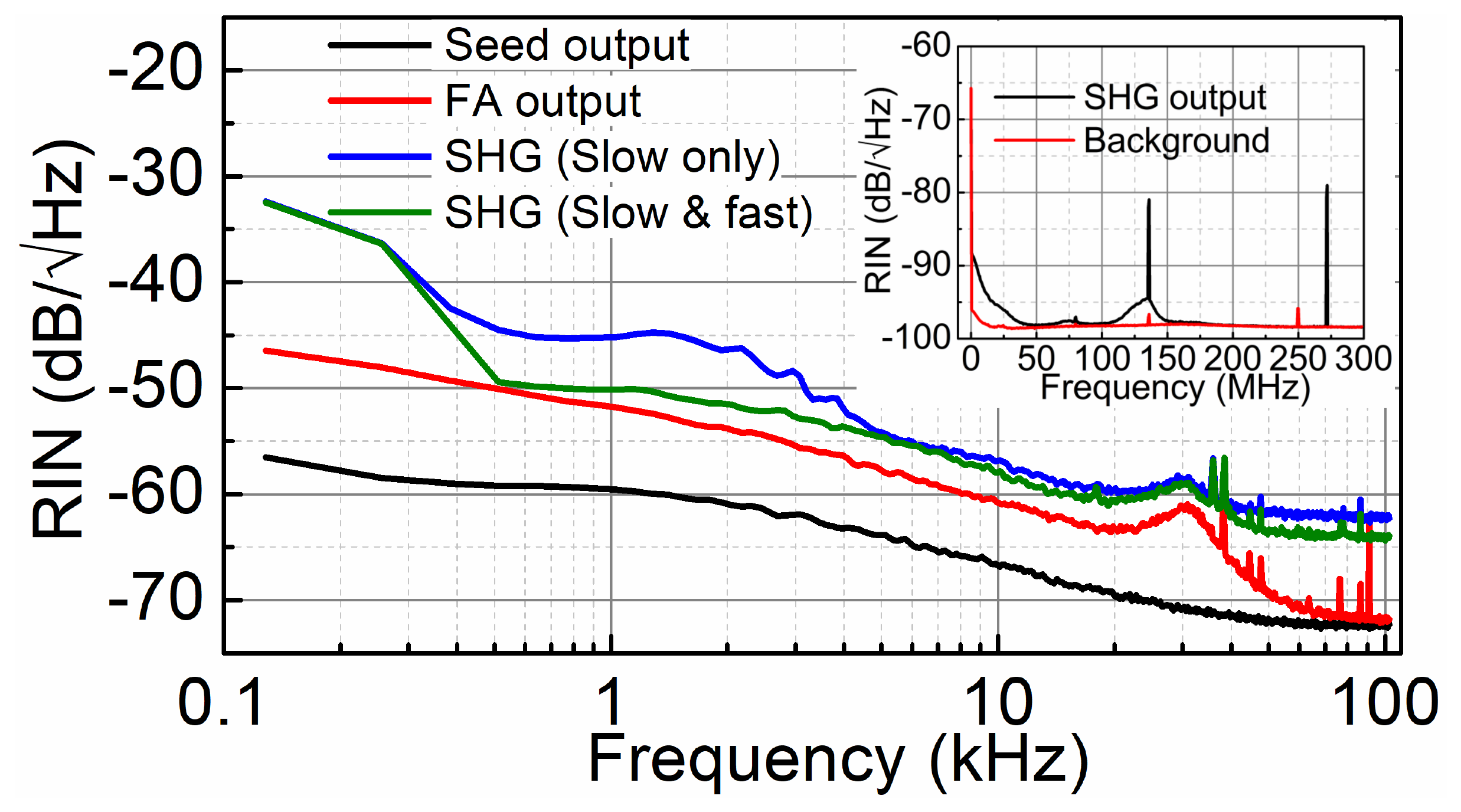}
    \caption{Relative Intensity Noise spectrum of the seed, fiber amplifier(FA), and SHG output from 128 Hz to 102.4 kHz. Inset shows the RIN at 0.1-300 MHz range, covering the rf frequency driving the fiber EOM.}
    \label{fig:RIN}
\end{figure}
Relative Intensity Noise(RIN) of the SHG output in continuous operation is shown in Fig. \ref{fig:RIN}.
We observe noisier RIN than the input light.
This is expected because mechanical noise also contributes to the intensity noise in the SHG output. 
The observed RIN also depended on the servo configuration.
Best RIN performance is observed when mechanical and laser modulation are both engaged.
Mechanical feedback alone left residual noise in the audio spectrum.
Residual amplitude and phase modulation of the fiber EOM caused weak RIN peaks at 130 MHz and 260 MHz.

\begin{figure}
    \centering
    \includegraphics[width=\columnwidth]{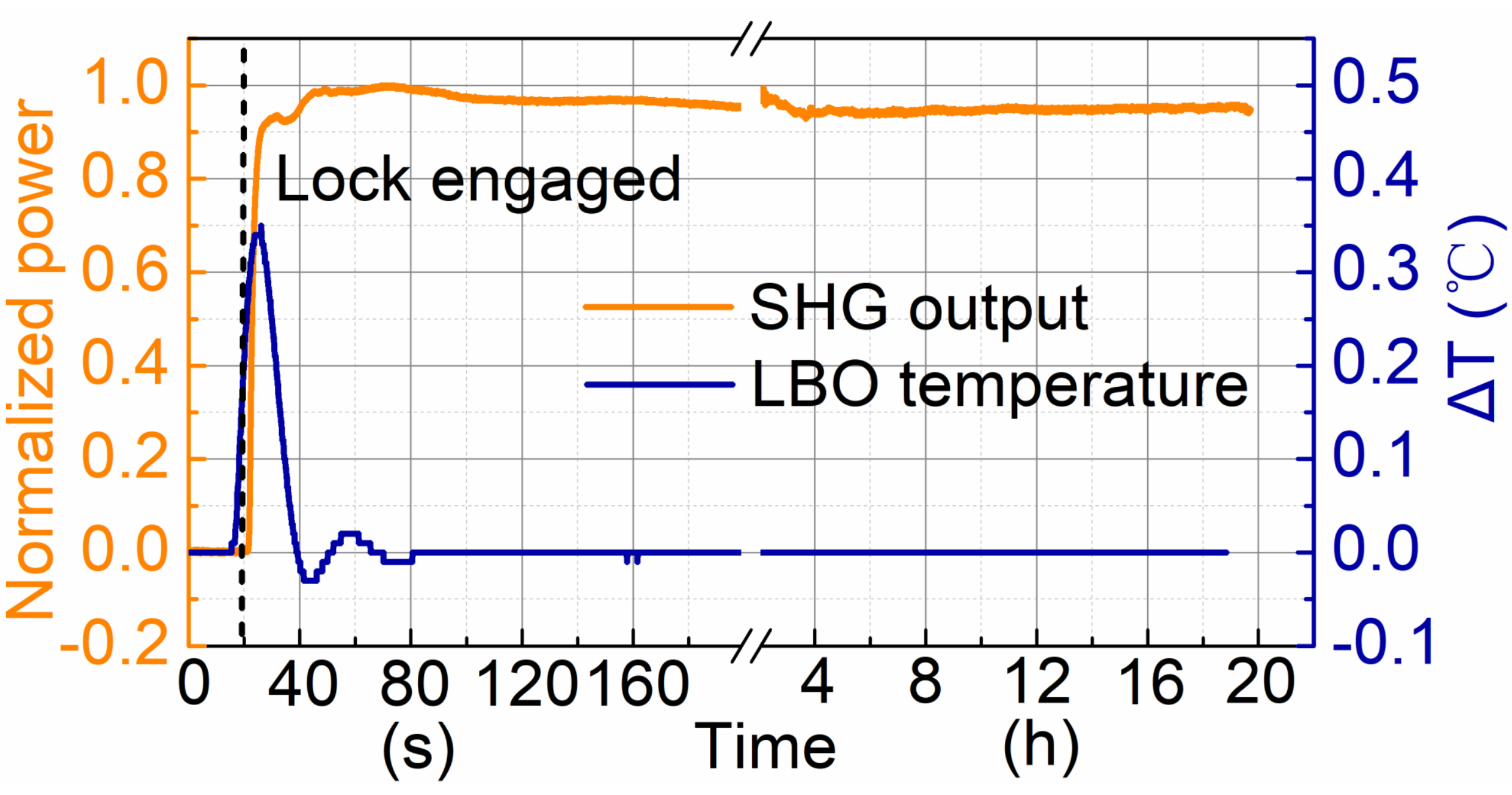}
    \caption{SHG output and LBO oven temperature from a cold-start. The laser is locked to the fundamental resonator mode at \(t=20\) s, and stayed locked during the acquisition window. Scale breaker indicates the crossover from transient to steady-state dynamics. SHG output is normalized by the observed maximum, 4.5 W, for this data.}
    \label{fig:power stability}
\end{figure}

SHG output and crystal temperature variation showing the long-term stability are presented in Fig. \ref{fig:power stability}.
When the lock engages, the oven temperature suddenly rises from absorption induced heating in the crystal at high circulating power.
This thermal perturbation settles down after a few minutes of active temperature control.
Sub-optimal choice of the oven set-point temperature caused the steady-state SHG output to be slightly lower than the true maximum.

\begin{figure}
    \centering
    \includegraphics[width=0.95\columnwidth]{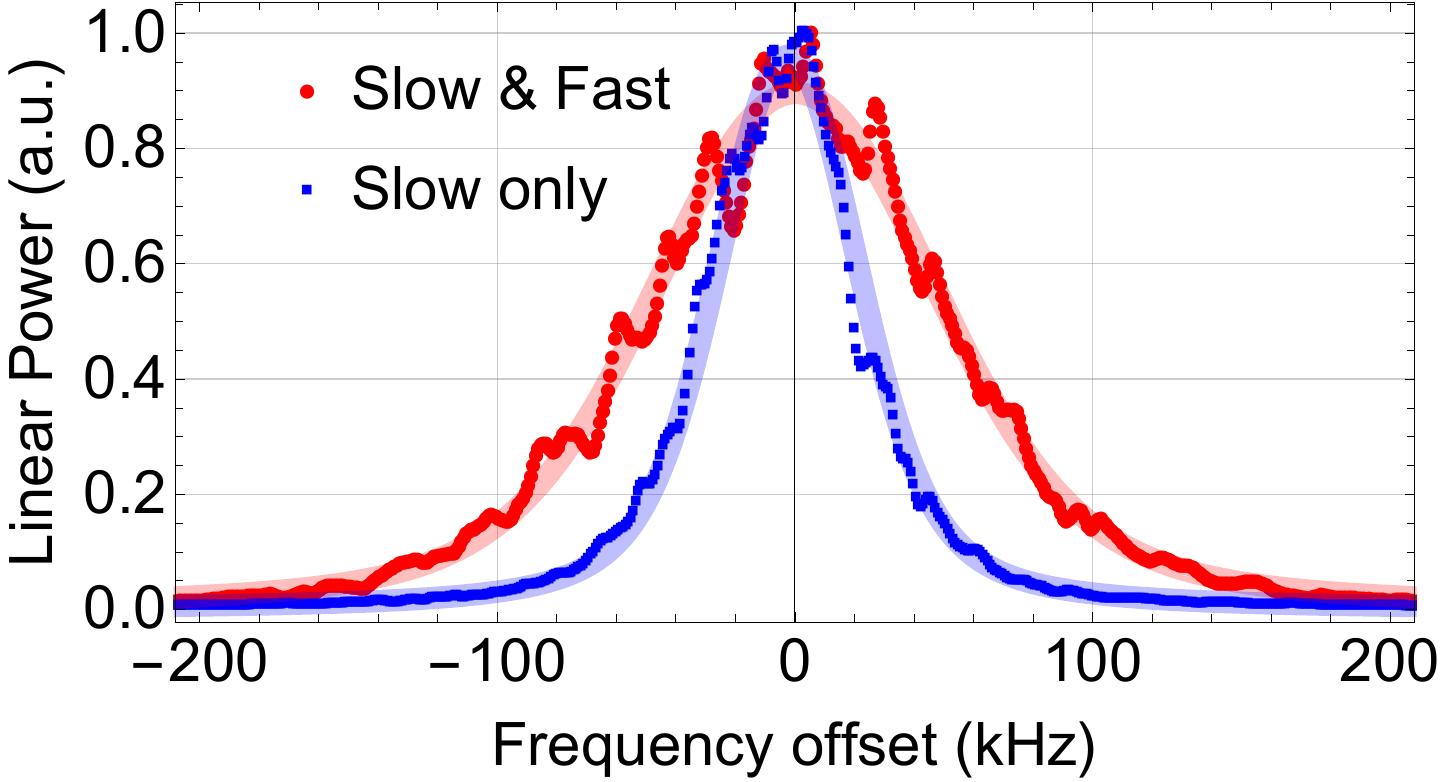}
    \caption{Self-heterodyne beatnote of the harmonic light using an 11 km long fiber delay line. Note that the FWHM of the self-heterodyne beatnote is twice the laser linewidth.  We observe FWHM/2 of 25(49) kHz without(with) the fast servo loop engaged.   
    The Voigt linewidths\cite{Chen2015c,Mercer1991}are 25(48) kHz for Slow only(Slow \& Fast servo loops). Measurements are averaged over \(100 \times \)10 ms sweeps(RBW 10 kHz).
    }
    \label{fig:SelfHeterodyne}
\end{figure}

The spectral quality of the harmonic light is characterized by a self-heterodyne measurement and is shown in Fig. \ref{fig:SelfHeterodyne}.
Extracted linewidths are 25 kHz for piezo stabilization only, and 49 kHz when the laser is additionally allowed to be modulated, which effectively broadens the linewidth.
The seed laser is specified to have  < 10 kHz linewidth, although we could not verify it due to spectral incompatibility of our self-heterodye setup.
The observed linewidths are not significantly broader then the seed.

\begin{figure}
    \centering
    \includegraphics[width=\columnwidth]{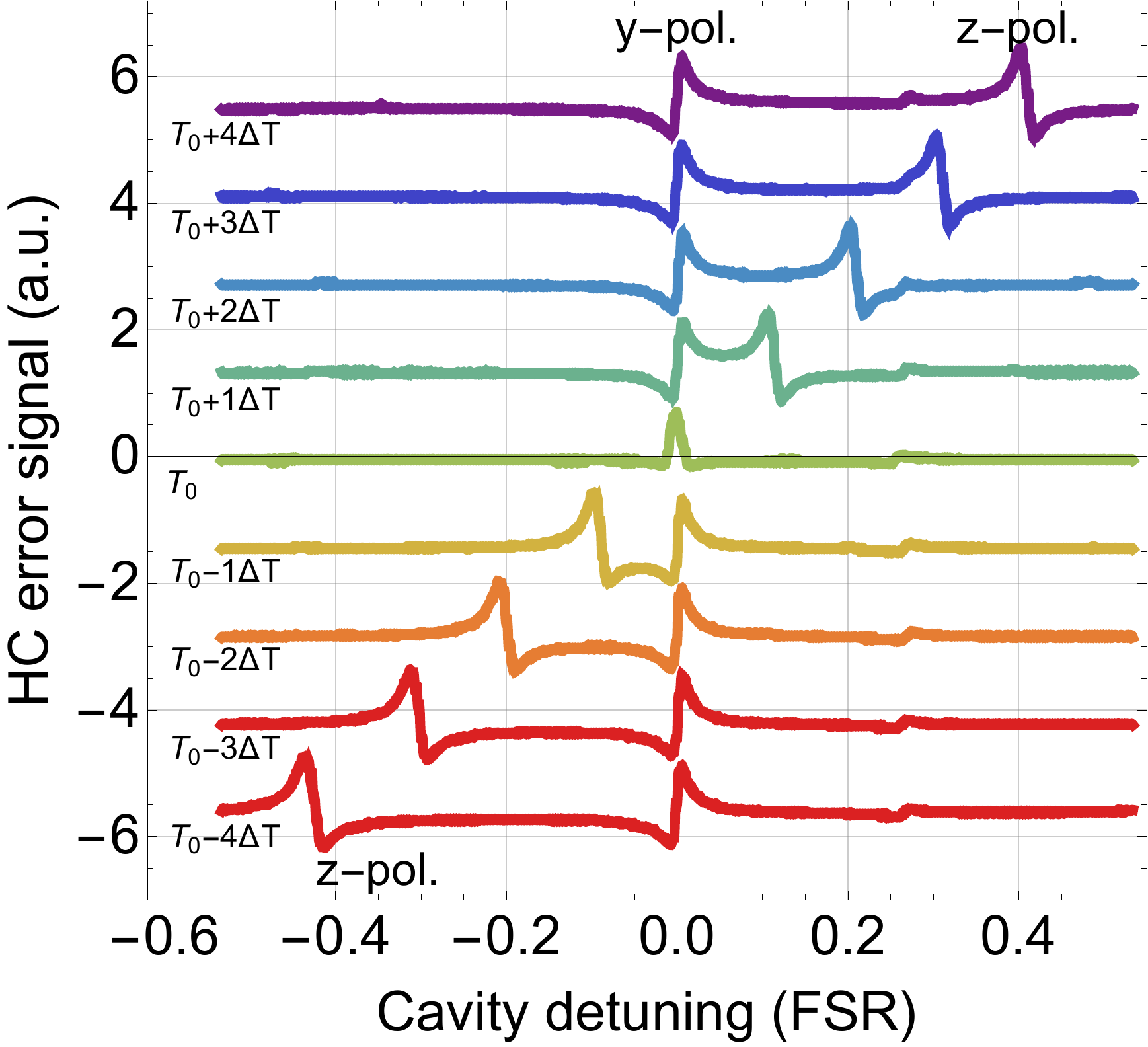}    \caption{Temperature-dependent HC error signal. Color represents different crystal temperatures, from \(80.57^{\circ}\)C to \(82.97^{\circ}\)C, in steps of \(\Delta \text{T}=+0.3^{\circ}\)C. Cavity detuning of zero corresponds to the \(y\)-pol. cavity resonance. \(y\)- and \(z\)-pol. become simultaneously resonant at \(T_{0}\), where the original error shape is lost.}
    \label{fig:HC error signal}
\end{figure}

Now we discuss the cavity stabilization in more detail.
Overall, the error signal generated by the PDH technique resulted in the best performance compared to two other techniques we tested.
The HC error signal could not withstand thermal transients, making it very difficult to cold start.
The HC error signal relies on intracavity birefringence, for our case \(n_{y}\neq n_{z}\).
The birefringence is also temperature-dependent(\(\frac{dn_{y}}{dT}\neq\frac{dn_{z}}{dT}\)), meaning that the indices of refraction for the two polarizations change at different rates.
As the temperature changes, their spectra in the cavity shift relative to each other, and so does the error signal. 
If the change is too large, they eventually become degenerate and the correct error signal shape is lost.
We directly observed the temperature-dependent HC error signal as shown in Fig. \ref{fig:HC error signal}.
When both polarizations become degenerate in the cavity, the slope of the error signal gets flipped and the servo lock is lost.
For our cavity, degenerate birefringence occurs if the optical path difference(OPD) between \(y-\) and \(z-\)polarizations become an integer multiple of \(\lambda=1540\) nm, where the OPD can be written as
\begin{equation}
    \text{OPD}_{\lambda}(T)=\left[n_{y}(T)-n_{z}(T)\right]\text{L}_{c}\left(1+\text{a}_{x} T\right).
    \label{eqn:OPD HC}
\end{equation}
Here, \(\text{L}_{c}\) is the length of the crystal, \(n_{y(z)}(T)\) are refractive indices of LBO for \(\lambda\) at the temperature T, and \(\text{a}_{x}\) is the coefficient of thermal expansion along the x-axis.
The exact condition for the degeneracy may vary depending upon the cavity configuration and spatial modes.
Based on the Sellmeier equations\cite{Kato1994}, at \(T_{0}=\)\(81.77^{\circ}\)C, we calculate the rate is \(\frac{d\text{OPD}}{dT}|_{T=T_{0}}\)=538 nm/\(^{\circ}\)C, suggesting that the HC lock cannot tolerate more than \(2.86^{\circ}\)C change.
This agrees to better than 2\% with the observed rate of 533 nm/\(^{\circ}\)C (or 0.346 FSR\(/^{\circ}\)C for our cavity).
We were able to analytically reproduce the observed distortion based upon a theoretical framework\cite{Marmet2011}.
Temperature-dependent non-linear loss changes the amplitude of the error signal slightly but the effect is minor.
In principle, one may consider slowly ramping the input power allowing the crystal to thermalize, minimizing temperature variation.
However, input-power dependent zero-crossing of the HC error signal requires in-situ adjustments and we found it to be impractical.
We concluded that the HC technique alone cannot be reliably used for this application.
A dither lock provided better stability than the HC lock, although its narrow capture range and dither frequency at tens of kHz caused significant intensity modulation, which can cause parametric heating\cite{Savard1997} of optically trapped atoms.

We believe that the limiting factor for more efficient conversion is linear loss associated with the crystal.
The LBO crystal exhibits two orders of magnitude larger photoacoustic absorption(\(y\)-pol., \(\sim10^{-3}~\text{cm}^{-1}\)) at \(1.5~ \mu\)m\cite{Waasem2013} than for \(\lambda \sim 0.5-1.1~\mu\)m.
Assuming the z-pol absorption is similar, a 30 mm long crystal will exhibit 0.3\% of absorption, which cannot be neglected.
Finite absorption combined with high circulating power causes substantial crystal heating and thermal effects.
These can lead to thermal lensing which will degrade the spatial mode quality and conversion efficiency at high power.
Using a loosely focused beam may mitigate the thermal lensing\cite{LeTargat2005a} which could make the HC technique a viable approach for stabilization.

In conclusion, we generated 14.0 W of cw single frequency, single transverse mode light at 770 nm, which is an unexplored wavelength for LBO at high power.
The achievable output power is mainly limited by the available fundamental power and residual absorption in the LBO crystal.

\textbf{Funding} National Science Foundation awards PHY-1104531, PHY-1521374, and the Air Force Office of Scientific Research Quantum Memories MURI. PY acknowledges support from the China Scholarship Council.



\end{document}